\documentclass{bioinfo}
\copyrightyear{2015} \pubyear{2015}
\access{Advance Access Publication Date: 20 Nov 2021}
\appnotes{Letter}
\usepackage{mathtools}
\usepackage{booktabs}
\usepackage{amsfonts}
\usepackage[para,online,flushleft]{threeparttable}
\usepackage{amssymb}
\usepackage{amsmath}
\usepackage{caption}
\usepackage{booktabs}
\usepackage{array}
\usepackage{paralist}
\usepackage{bigstrut}
\usepackage{multirow}
\usepackage{dcolumn}
\usepackage{siunitx}
\usepackage{amsmath}
\usepackage{tikz}
\usetikzlibrary{automata,arrows,positioning,calc}
\usepackage{graphicx}
\usepackage{tabularx}
\newcolumntype{Y}{>{\centering\arraybackslash}X}

\begin{document}
\firstpage{1}

\subtitle{Genome Analysis}

\title[CTMC in evolution]{Continuous-time Markov chain as a generic trait-based evolutionary model}
\author[Amiryousefi, A]{Ali Amiryousefi\,$^{\text{\sfb 1}*}$}
\address{$^{\text{\sf 1}}$Systems Oncology Unit, Faculty of Medicine, University of Helsinki, Helsinki, Finland.}

\corresp{$^\ast$To whom correspondence should be addressed.}

\history{Received on X; revised on X; accepted on X}

\editor{Associate Editor: X} 

\abstract{More than ever, today we are left with the abundance of molecular data outpaced by the advancements of the phylogenomic methods. Especially in the case of presence of many genes over a set of species under the phylogeny question, more sophisticated methods than the crude way of concatenation is needed. In this letter, by placing the continuous-time Markov chain (CTMC) on the species set, I present a novel model for inferring the phylogeny, obtaining the network graph, or drawing the proximity conclusions. The rate of transition between states is calculated based on the binary character paths between each two species. This is the base for the pairwise distances between species. Next to its generic use, the formulation of the model allows the site-wise phylogenetic inference and a mathematically justified method of combining these information to form as big as the whole genome phylogenetic inference. Although based on the characters or traits, this model is inherently a distance method but its advantage to other methods of the same class is its ability to incorporate the information of all the other species in forming the pairwise distance between two them.
\textbf{Contact:} \href{ali.amiryousefi@helsinki.fi}{ali.amiryousefi@helsinki.fi}
}

\maketitle

\section*{B{\sc ACKGROUND}}
Given a set of species, there are many methods to call for their phylogenetic tree based on their specific characteristics. In specific, in the molecular evolution context, the richness of the methods both in the quantity and quality is significant. This richness should not indicate the fallacy of the exactness of the models since each model makes its own simplifications and assumptions with regard to the evolutionary mechanism. For example, the distance based methods base their distances on nucleotide evolutionary model that often controversy arise for the choice of an appropriate one. The character based methods, besides this dilemma, carry on the burden of the sites independence and furthermore, likelihood based models suffers from an expensive computational efficiency in finding the right tree. Molecular clock is another assumption that most of the models explicitly make. On the other hand, different models, use the available data with a different  degree of efficiency. The parsimony based methods for example, will not use the sites with at most one different characters and distance based methods will neglect the sites with gaps in their analysis. Nonetheless, these models were able to give us reasonably good phylogenies with some statistical properties that made them readily more interesting choices to form the phylogenetic tree when the molecular data is available \citep{wheeler2012systematics, felsenstein2004inferring}. 

Here with placing the CTMC over the species set rather than character set, I introduce a novel point of view toward evolution of a molecular sequences. This minimal modification to the distance methods, will leave us with a significant rewards in delineating the genealogical relationship of a given set of species. The rate of transition between states of the chain will be calculated based on the frequency of the character paths between each two states. The embedded transition matrix will then be calculated for each site and the multiplication of these matrices over all sites will be the final transition matrix that is viewed as the species distances. The advantage of this model is its ability to sum over all the other observed characters of species when forming the distance between two species. This is dramatically different from the basic distance methods that after adhering to a specific evolutionary model, will calculate the pairwise distance of two sequences irrespective of the other species distances. As long as the goal of any phylogenetic method is to decipher the historical dependency of the species, this implicit ignorance, implying the independence of species is abominable. The representation of the model enables a mathematically justified method of combining different evolutionary information among the species and a computationally tractable solution of the phylogenetic tree in the presence of massive data. The former property is specially attractive when interested in the species tree with the availability of more than one gene tree or alignment while latter is appealing in the existence of the genomic size data over species set. The other advantage of this formulation is its ability to integrate the sites with gaps and its mixing ability that allows elucidation of a single phylogenetic tree in the existence of different types of independent evolutionary characters throughout species.

\section*{N{\sc EIGHBORING} B{\sc OX} P{\sc ROCESS}}
Here I define the neighbouring box process (NBP), as the generative model of the generic trait evolution.
Consider the evolution of an artificial species in a box with only one trait. This trait can be a letter from any closed alphabet set. Furthermore consider this trait to be governed by an evolution machinery under an assumed molecular clock. This indicates the passage of exponentially distributed amount of time before any substitution or deletion takes place. The emergence of a new species through this species therefore can be considered as an exponential distribution too, albeit with different parameters. Now lets assume that the box is embedded with a speciation and extinction switches. Anytime the species in the box hits the speciation switch, a new artificial species will be formed in a neighboring brand new box and start its evolutionary life like its anscestor species. On the other hand, each time the species hits the extinction switch, the box in which it is living, will be vanished into abyss\footnote{The case of insertion can be considered as the incidence of new trait that can be put in a new box ruled by the same molecular mechanism}. The state of the species at anytime is the letter observed for that trait. This is a process that is assumed to be followed along the time from past till now and we are given a number of boxes with their corresponding observed letters of a trait. From the time of emergence of a species the fate of the evolution under this process can be depicted. Only one of the realization of this process has happened in the past time leaving us with the diversity of species around us and the goal of each evolutionary model is to estimate this specific line of history as closely as possible. In the following section I introduce one of such models based on the CTMC which imitates the latter process assumptions.

\section*{T{\sc HE} M{\sc ODEL} F{\sc ORMULATION}}

Assume that a neighboring box process with an initial species is started at some point in the past given rise to number of species at this time. For a specific trait, each species is registered with a letter and this is all the information that we have about the evolution history of the set of species in hand. We can assume that all the species in each box can communicate with each other with a given rate. This is like assuming that the species are the states of an ergodic CTMC which are communicating through the two way paths represented by the letters of two states. Consider the vector $\textbf{v}=(v_1, v_2, \ldots, v_n)$ as the site characters of a given alignment for $n$ species $\textbf{s}=(s_1, s_2, \ldots, s_n)$. These characters can be a letter of any known closed alphabet set. In specific, these sets can be any of $\mathcal{N}=\{A,C,G,T\}$, $\mathcal{A}=\{His,Asp, \ldots, Val\}$, $\mathcal{C}=\{UUU,UUC, \ldots ,GGG\}$, or $\mathcal{M}=\{1,2,\ldots\}$ as nucleotide, amino acid residues, codons, or morphological attributes, respectively.\\ 
Given this vector, we define the path matrix of $\textbf{V}= \textbf{v}\otimes\textbf{v}.$ This is the matrix which shows all the possible alphabetic paths from each species to another. 
\begin{equation} 
V_{n,n} = \mathbf{v}\mathbf{v}^T=
 \begin{pmatrix}
  v_{1}v_1 & v_{1}v_2 & \cdots & v_{1}v_n \\
  v_{2}v_1 & v_{2}v_2 & \cdots & a_{2}v_n \\
  \vdots  & \vdots  & \ddots & \vdots  \\
  v_{n}v_1 & v_{n}v_2 & \cdots & v_{n}v_n 
 \end{pmatrix}.
\end{equation}
Now each element of this matrix shows an alphabetic path through which two species communicate. These communications can form a saturated network of species. The species can be considered as the states of a CTMC and we can use the frequency proportion of the paths in the network as the transition rate between species. With defining $f_{v_.}$ as the frequency of $v_.$ letter in the site, one can obtain the transition rate matrix $Q_{n,n}$ based on the collection of all these values as follow,
\begin{equation}
Q_{n,n}=(q_{ij})=
	\begin{cases}
     \frac{f_{v_i} f_{v_j}}{n(n-1)} & \quad \text{if } i\neq j , v_i\neq v_j\\
     \frac{f_{v_i}(f_{v_i} -1) }{n(n-1)} & \quad \text{if }	i\neq j , v_i=v_j\\
      -\sum_{i\neq j} q_{ij}       & \quad \text{if } i=j \\
  \end{cases}.
\end{equation}
Upon obtaining the transition rate matrix $Q$, we can derive the embedded stochastic transition matrix $P$ as,

\begin{equation}
P_{n,n}=(p_{ij})=
	\begin{cases}
     \frac{q_{ij}}{\sum_{i\neq j} q_{ij}} & \quad \text{if } i\neq j\\
       1-\sum_{i\neq j} p_{ij}   & \quad \text{if } i=j \\
  \end{cases}.
\end{equation}
Ideally, we are more interested on the transition matrix $P_t$ so that we can answer more queries with respect of time variable in Markov process. On the other hand, the calculation of this matrix is computationally expensive and even intractable in the cases of many species. This factor, beside the long enough time in history to permit for the set of species to emerge, renders the use of infinite time value an interesting choice\footnote{Note that if one could derive the $P_t$ then setting the $t$ as the entropy of $\mathbf{v}$ would be a reasonable choice so that the least conserved sites have the most time to be emerged. This choice of $t$ is quite natural since if there is higher entropy in the site this represents a larger scale of time for the evolution of the neighboring box process. Upon this imputation of time variable in transition matrix, we will be left with a stochastic matrix $P$.}. Since this matrix is denoting an ergodic Markov chain. Furthermore, through the formulation of $Q$, it follows that the Markov chain is reversible. For example a simple scenario of 4 species with $\mathbf{v}=(x,x,y,z)$ is leading to the following network graph of this site as well as the binary path matrix $V_{4,4}$;
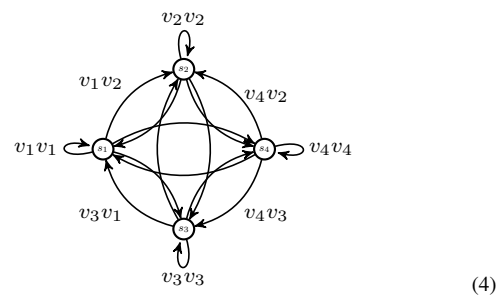
\begin{equation}\label{graph}
\begin{tikzpicture}[->, >=stealth', auto, semithick, node distance=3cm]
\tikzstyle{every state}=[fill=white,draw=black,thick,text=black,scale=0.5]
\node[state]    (A)                     {$s_1$};
\node[state]    (B)[above right of=A]   {$s_2$};
\node[state]    (C)[below right of=A]   {$s_3$};
\node[state]    (D)[below right of=B]   {$s_4$};
\path
(A) edge[loop left]     node{$v_1v_1$}     (A)
    edge[bend left]     node{$v_1v_2$}     (B)
    edge[bend left,below]        (D)
    edge[bend left]         (C)
(B) edge[loop above]		node{$v_2v_2$}	(B)
	edge[bend left]			(C)
	edge[bend right]          (D)
	edge[bend left]			(A)
(C) edge[loop below]		node{$v_3v_3$}		(C)
	edge[bend left]       (D)
	edge[bend left]		(B)
	edge[bend left]	node{$v_3v_1$}			(A)
(D) edge[loop right]    node{$v_4v_4$}     (D)
    edge[bend right,right]    node{$v_4v_2$}      (B)
    edge[bend left]     node{$v_4v_3$}      (C)
    edge[bend left,above]             (A);
\end{tikzpicture}
\end{equation}

\begin{equation} 
V_{4,4} = 
 \begin{pmatrix}
  xx & xx & xy & xz \\
  xx & xx & xy & xz \\
  yx  & yx  & yy & yz  \\
  zx & zx & zy & zz 
 \end{pmatrix}.
\end{equation}

\section*{C{\sc ONCLUSION}}
While relaxing the choice of evolutionary model, the neighboring box process is still preserving the site independence and molecular clock assumption often tied with other evolutionary models. Furthermore,  considering the infinite time for the evolution and since $P^{(t)}=e^{Qt}$, we have $P^{(\infty)}=P$. Gathering all these matrices for each species in our set, we can form the distance matrix $D= \frac{1}{P_1 \times P_2 \times \ldots P_n}$ as the basis of tree or network derivation of the given set of species.  





\bibliographystyle{natbib}
\bibliography{Document} 









\end{document}